\documentclass[aps, twocolumn, showpacs, amsmath]{revtex4}
\usepackage{dcolumn}
\usepackage{bm}
\usepackage{graphicx}
\usepackage{color}
\usepackage{subfigure}
\usepackage{hyperref}
\usepackage{latexsym}
\usepackage{amsthm}
\usepackage{amssymb}

\DeclareGraphicsExtensions{.jpg,.pdf, .mps, .png, .eps, .ps, .EPS,.gif}

\begin{document}
\def\be{\begin{equation}}
\def\ee{\end{equation}}

\def\bc{\begin{center}}
\def\ec{\end{center}}
\def\bea{\begin{eqnarray}}
\def\eea{\end{eqnarray}}
\newcommand{\avg}[1]{\langle{#1}\rangle}
\newcommand{\Avg}[1]{\left\langle{#1}\right\rangle}

\def\ie{\textit{i.e.}}
\def\etal{\textit{et al.}}
\def\m{\vec{m}}
\def\G{\mathcal{G}}

\title{ Mutually connected component of network of networks with replica nodes}
\author{  Ginestra Bianconi}

\affiliation{
School of Mathematical Sciences, Queen Mary University of London, London, E1 4NS, United Kingdom}

\author{ Sergey N. Dorogovtsev}
\affiliation{Departamento de F\'{\i}sica da Universidade de Aveiro $\&$ I3N, 3810-193, Aveiro, Portugal} 
\affiliation{Ioffe Physico-Technical Institute, 194021 St. Petersburg, Russia}

\author{Jos\'e F. F. Mendes}
\affiliation{Departamento de F\'{\i}sica da Universidade de Aveiro $\&$ I3N, 3810-193, Aveiro, Portugal}

\begin{abstract}
We describe the emergence of the  giant mutually connected component in  networks of networks in which each node has a single replica node in any layer and can be  interdependent only on its replica nodes in the interdependent layers. 
We prove that if 
in these networks, 
all the nodes of one network (layer) are interdependent on the nodes of the same other interconnected 
layer, then, remarkably, 
the mutually connected component does not depend on the topology of the network of networks. 
This component 
coincides with the mutual component of 
the fully connected network of networks constructed from the same set of layers, i.e., a multiplex network. 
\end{abstract}
\maketitle

\section{Introduction}

Complex networks structures strongly affect 
cooperative and critical phenomena 
in them \cite{Dynamics,crit}. Despite the huge interest in the topic only recently it has become clear that in order to characterize the function and the dynamics of the majority of complex systems, it is necessary to make a step further and consider  networks of networks \cite{Boccaletti_rev,Arenas_rev}. For example, if we want to understand the robustness of critical infrastructures \cite{Havlin1} it is necessary to characterize the complex interdependencies between them, or if we aim at characterizing the function of a cell, we need to scale up the analysis of single cellular networks such as 
a protein interaction network and 
a metabolic network and study also 
interactions between different cellular networks.
Critical phenomena 
in a network of networks and multilayer structures \cite{Boccaletti_rev,Arenas_rev,PRE} show surprising new features \cite{Havlin1,Havlin2,Son,Dorogovstev,Goh,Kabashima, JSTAT,Diffusion, Boguna, dedomenico,Gao:gbhs2011,Gao:gbhs2012,Havlin3,Gao:gbsxh2013,Dong:dtdxzs2013,Dong:dgdtsh2013,BianDoro,BD2}.
In particular it has been recently shown \cite{Havlin1, Havlin2,Son,Dorogovstev} that when we consider several interdependent networks, the system as a whole might be much more fragile than single networks taken in isolation, and that the interdependencies between different networks can trigger cascading failure events of dramatic impact on the networks of networks. 

After the seminal work \cite{Havlin1} it has become commonly accepted that the robustness of interdependent networks can be evaluated by considering the size of the mutually connected component of the interdependent networks.
While the emergence of the giant mutually connected component in multiplex networks (graphs with nodes of one kind and different types of links) is already well understood \cite{Havlin1,Havlin2,Son,Dorogovstev, Goh,Kabashima}, few works focused on 
the mutually connected component in a network of networks 
in which nodes in each individual network (layer) are interdependent with nodes in some other layers \cite{Gao:gbhs2011,Gao:gbhs2012,Havlin3,Gao:gbsxh2013,Dong:dtdxzs2013,Dong:dgdtsh2013,BianDoro}. 
In the interdependent networks, a mutually connected component is introduced as a subgraph remaining after the cascade of failures spreading 
back-and-forth 
through intralinks within layers and through 
interlinks---interdependencies---from layer to layer.  
For a sufficiently large number of layers in the network of networks \cite{Gao:gbhs2011}, say, greater than $3$, the complex structure of interconnections makes the problem of the mutually connected component 
principally richer than for a pair of 
interdependent  
networks \cite{Havlin1,Havlin2,Son}.
The key question is how far are complex networks of networks from the multiplex networks in respect of their mutual components?
In a series of publications \cite{Gao:gbhs2011,Gao:gbhs2012,Havlin3} a network of networks in which each node can be linked to a random node in an interdependent layer (i.e., there are no replica nodes) was considered. It was demonstrated that if this type of  network of networks is a proper tree (in respect of 
interlinks), then the size of its mutual component is 
determined by the number of layers and does not depend on the structure of this tree; but it was found, contrastingly, that  
for a network of networks with loops, 
its global structure matters
\cite{Gao:gbsxh2013}.  
In the works \cite{Gao:gbhs2011,Gao:gbhs2012,Havlin3,Gao:gbsxh2013} it was found that  if the supernetwork of interdependencies between the layers is a tree, then the mutually connected component  of such a network of networks follows the same equations as for a multiplex network formed by these layers. 
In contrast, it was shown in Ref.~\cite{Gao:gbsxh2013} that the supernetwork of interconnections in a network of networks of this class (each node in a layer is interconnected with a random node in a respective interdependent layer) contains loops, then the mutually connected component does not satisfy these equations. More recently, it has been found in Ref.~\cite{BD2} that the difference between these topologies can be dramatic. While in the case of a tree supernetwork, all the layers percolate at once resulting in a single discontinuous phase transition, in the case of a loopy supernetwork with heterogeneous degrees of the nodes, layers with different number of interdependencies fail one after the other 
as the initial damage inflicted to the network increases 
\cite{BD2}. This results in a chain of phase transitions. 

\begin{figure}
\begin{center}
{\includegraphics[width=3.1in]{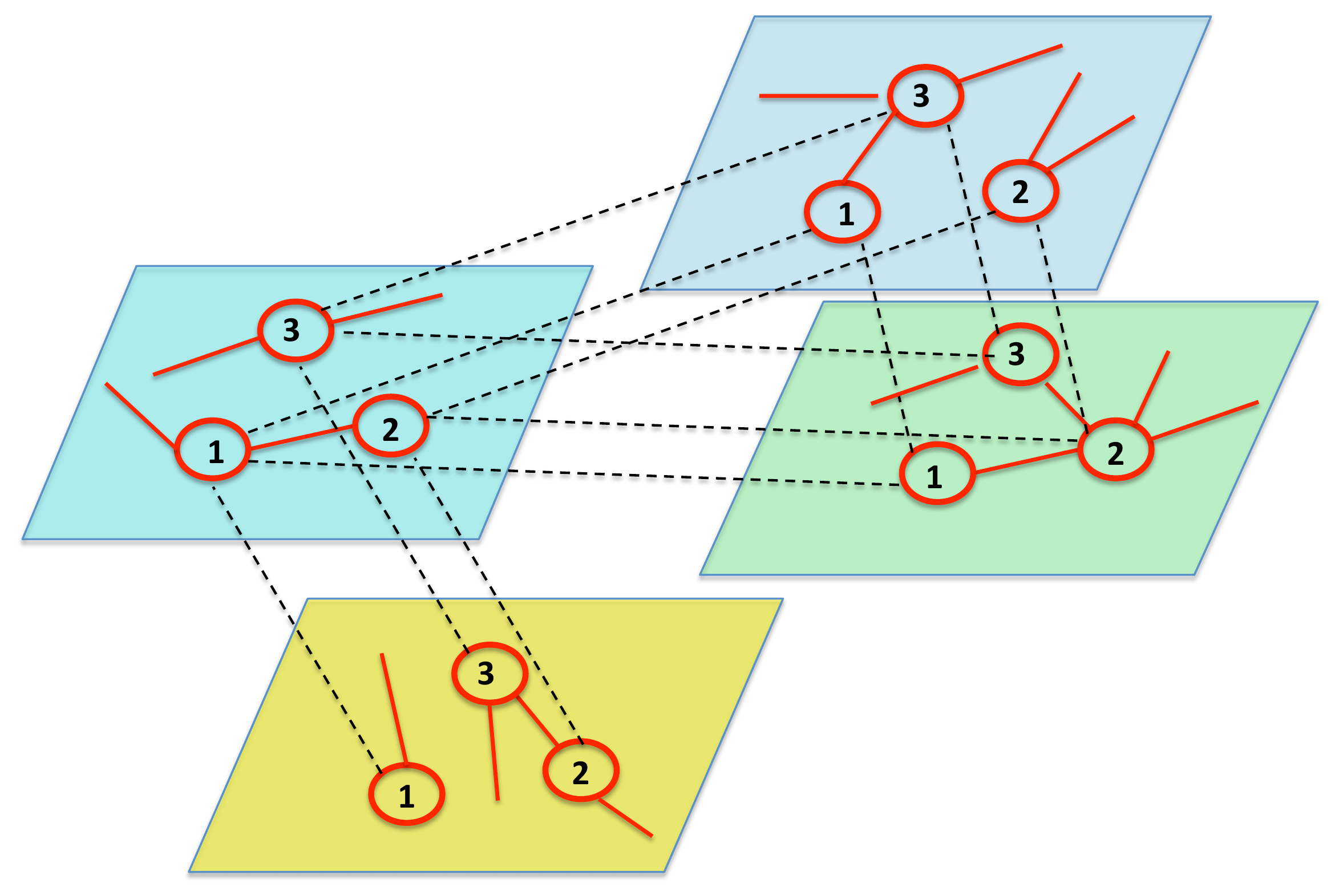}}
\end{center}
\caption{(Color online)
Schematic view of a typical network of networks with replica nodes considered in this paper. Interdependencies  
(interlinks 
between nodes from different levels) are shown by the black dashed lines. Intralinks between nodes within layers are shown as solid red lines.
} 
\label{f1}
\end{figure}

Here  we will show  that the way the nodes are connected between the layers affects very significantly the robustness properties of the network of networks. In fact  if each node has a single replica nodes in the other layers and  the interconnections are organized in such a way that all the nodes of 
a layer are interdependent on the nodes of the same other layer (see Fig.~\ref{f1}), then, 
remarkably, 
the mutually connected component  
does not depend on topology of the network of networks. 
We emphasize that it can be tree or it can be loopy, in both cases, apart from natural dependence on the structures of individual layers, only the size of the network of networks (the number of layers) matters. 
Thus we show that the problem of a wide range of networks of networks with replica nodes is actually reduced to the well studied problem of multiplex networks. 
We will arrive at this unexpected conclusion by applying the convenient message-passing technique \cite{Son,Kabashima,JSTAT,Mezard,Weigt,karrer2014percolation}. In the case of locally tree-like networks, this approach, which is also called the cavity method, believe propagation, message-passing equations, etc., leads to the same equations as the more traditional technique \cite{newman2001random} used in Refs.~\cite{Havlin1,Son, Dorogovstev, Havlin3,Gao:gbhs2012,Gao:gbhs2011,Gao:gbsxh2013} and in numerous works for other complex networks \cite{crit}. On the other hand, when a network has finite loops, the message passing technique provides an approximate solution, which was reported to be reasonably precise in investigated loopy networks \cite{karrer2014percolation}. In addition, this powerful techniques is particularly convenient for formalization and unified consideration of the class of problems under consideration, 
and so it constitutes a useful framework for analysis of a wide range of these complex networks. 

\section{ Network of networks with replica nodes}
A network of networks 
in this work 
is formed by $M$ networks $\alpha=1,2,\ldots, M$ each 
of $N$  nodes $i=1,2,\ldots, N$. 
We assume that $M$ is finite and $N$ is infinite. 
Every node $(i,\alpha)$ can be connected to nodes $(j,\alpha)$ in within the same network or with its ``replica nodes'' $(i,\beta)$ in other networks.
If 
network $\alpha$ is interdependent with 
network $\beta$, each node $(i,\alpha)$ of network $\alpha$ is interdependent on node $(i,\beta)$ of network $\beta$ and vice versa (see Fig.~\ref{f1}). We define the network of networks with a super-adjacency matrix of elements $a_{i\alpha,j\beta}=1$
if there is a link between node $(i,\alpha)$ and node $(j,\beta)$ and zero otherwise. In these 
specific 
networks,  
$a_{i\alpha,j\beta}=0$
if both $i\neq j$ and $\alpha\neq \beta$. 
We call the graph of interdependencies with an adjacency matrix $A_{\alpha\beta}$, such that for every $i$ $A_{\alpha\beta}=a_{i\alpha,i\beta}$, the supernetwork, ${\cal G}$, of the network of networks. 

The mutually connected component can be defined as follows.
Each node $(i,\alpha)$ is in the mutually connected component if it has at least one neighbor $(j,\alpha)$ which belongs to the mutually connected component and if all the linked nodes $(i,\beta)$ in the interdependent networks are also in the mutually connected component.

From this definition, it follows that iff a node $i$ in one layer of a connected (in terms of interlinks) network of networks is in a mutual component, then all its ``replica'' nodes in all layers are in this mutual component. Then, for infinite $N$, the same set of nodes $\{i\}$ in each level belong to the giant mutually connected component. 
Moreover, it is clear from the same definition that the set of replicas $\{i\}$ belonging to the mutual component will not change if we rearrange connections in the supernetwork retaining it connected. 
A cascade of failures in the networks of networks can be initiated by deleting a fraction of nodes (as is natural, this removal keeps all interdependencies functioning). 
After such a removal of a node, all its replica nodes necessarily should fall apart from the mutual component, so our conclusions do not change.  
In this paper we will prove strictly that these 
heuristic arguments 
are correct.

For a given network of networks it is easy to construct a message passing algorithm that allow us to determine if node $(i,\alpha)$ is in the mutually connected component. 
We denote by $\sigma_{i\alpha \to j\alpha}=1,0$
the  message in within a layer, from node $(i,\alpha)$ to node $(j,\alpha)$ and indicating $\sigma_{i\alpha \to j\alpha}=1$ 
if node $(i,\alpha)$ is in the mutually connected component when we consider the cavity graph by removing the link $(i,j)$ in network $\alpha$. 
Furthermore, 
let us denote by  
$S'_{i\alpha \to i\beta}=0,1$
the  message between the ``replicas'' $(i,\alpha)$ and $(i,\beta)$ of node $i$ in layers $\alpha$ and $\beta$. The message  
$S'_{i\alpha \to i\beta}=1$
indicates if the node $(i,\alpha)$ is in the mutually connected component when we consider the cavity graph by removing the link between node $(i,\alpha)$ and node $(i,\beta)$.
 In addition, 
we indicate with  
$s_{i\alpha}=0$ 
a node that is removed from the network as an effect of the damage inflicted to the network, otherwise  
$s_{i\alpha}=1$. 

The message passing equations for these messages are of the following forms:
\bea
&&\!\!\!\!\!\!\!\!\!\!\!\!\!\!\!
\sigma_{i\alpha\to j\alpha} = s_{i\alpha}\!\!\!\!\prod_{\beta\in {\cal N}(\alpha)}\!\!\!\!S'_{i\beta \to i\alpha}
\!\left[1-\!\!\!\!\prod_{\ell\in N_{\alpha}(i)\setminus j}\!\!\!\!(1-\sigma_{\ell\alpha \to i\alpha})\right],
\nonumber 
\\[5pt]
&&\!\!\!\!\!\!\!\!\!\!\!\!\!\!\!
S'_{i\alpha \to i\beta} = s_{i\alpha}\!\!\!\!\prod_{\gamma\in {\cal N}(\alpha)\setminus \beta}\!\!\!\!S'_{i\gamma \to i\alpha}
\!\left[1-\!\!\!\!\prod_{\ell\in N_{\alpha}(i)}\!\!\!\!(1-\sigma_{\ell\alpha \to i\alpha})\right],
\label{mp1}
\eea
where $N_{\alpha}(i)$ indicates the set of nodes $(\ell,\alpha)$ which are neighbors of node $i$ in network $\alpha$, and ${\cal N}(\alpha)$ indicates the layers  that are interdependent on network $\alpha$.
Using simple properties of these messages, the expression for the messages $\sigma_{i\alpha\to i\beta}$ can be simplified giving
\bea
&&\!\!\!\!\!\!\!\!\!\!\!
\sigma_{i\alpha \to j\alpha} = S'_{i\alpha\to i\beta}S'_{i\beta \to i\alpha}
\!\left[1-\!\!\!\!\prod_{\ell\in N_{\alpha}(i)\setminus j}\!\!\!\!(1-\sigma_{\ell\alpha \to i\alpha})\right],
\label{sim}
\eea
(see Appendix  \ref{ApA} for the detailed derivation).
Finally 
$S_{i\alpha}$
indicates if   a node $(i,\alpha)$ is in the mutually connected 
component or not 
($S_{i\alpha}=1,0$), 
namely 
\be
S_{i\alpha} = s_{i\alpha}\!\!\!\!\prod_{\beta\in {\cal N}(\alpha)}\!\!\!\!S'_{i\beta \to i\alpha}
\left[1-\!\!\!\!\prod_{\ell\in N_{\alpha}(i)}\!\!\!\!(1-\sigma_{\ell\alpha \to i\alpha})\right],
\label{S}
\ee


\section{ General solution of the algorithm}
\label{par3}

In this paragraph we will show that the general solution of the message passing algorithm in Eqs.~(\ref{mp1}) in a network of networks with replica nodes, where the supernetwork may consist of an arbitrary number of connected components, is given by 
\bea
&&\hspace*{-10pt}
\sigma_{i\alpha \to j\alpha} = 
\prod_{\beta\in {\cal C}({\alpha})\setminus \alpha}\left\{s_{i\beta}\left[1-\prod_{\ell\in N_{\beta}(i)}(1-\sigma_{\ell\beta \to i\beta})\right]\right\}
\nonumber \\[5pt]
&&\hspace*{24pt}
\times s_{i\alpha}\left[1-\prod_{\ell\in N_{\alpha}(i)\setminus j}(1-\sigma_{\ell\alpha \to i\alpha})\right],
\label{mesnew} \\[5pt]
&&\hspace*{-9pt}
S_{i\alpha} = \prod_{\beta\in {\cal C}(\alpha)} \left\{\!s_{i\beta}\!\! \left[1-\prod_{\ell\in N_{\beta}(i)}(1-\sigma_{\ell\beta \to i\beta})\right]\!\!\right\}
. 
\label{Smesnew}
\eea 
 Here and in the following ${\cal C}(\alpha)$ is the connected component in the supernetwork ${\cal G}$ to which  layer $\alpha$ belongs. In particular, if the supernetwork contains only a single connected component, then ${\cal C}(\alpha)={\cal G}$ for every layer $\alpha$ and the cardinality of this component (its size) $|{\cal C}(\alpha)|$ coincides with $M$, the total number of layer of the network of networks. 

In the Appendix ${\ref{ApB}}$ we show that indeed this formula is valid for simple examples of supernetworks, such as a tree, a forest and a single loop.

Here in the following we want to show that Eqs.~(\ref{mesnew}) and (\ref{Smesnew}) are valid for every supernetwork topology.
If we consider a connected supernetwork, Eq.~(\ref{mp1}) for the messages 
$S'_{i\alpha \to i\beta}$
can be written as
\bea
&&\hspace*{-11pt}
S'_{i\alpha \to i\beta} = 
\!\!\!\!\prod_{{\cal P}_{\alpha\beta}(\gamma)}\!\! 
\prod_{\begin{array}{c}\phantom{.}\\[-26pt]\\\,\scriptstyle{\gamma'}\in {\cal P}_{\alpha\beta}(\gamma)\\[-3pt] \mbox{\scriptsize{with}}\\[-5pt]
\scriptstyle{a_{i\gamma,i\gamma'}=1}\end{array}}\!\!\!\!\!\!\!\!\!\!
S'_{i\gamma \to i\gamma'}\!
\left\{\!s_{i\alpha}\!\!\left[1\!-\!\!\!\!\!\prod_{\ell\in N_{\alpha}(i)}\!\!\!\!(1{-}\sigma_{\ell\alpha \to i\alpha})\right]\!\!\right\}
\nonumber 
\\
&&\hspace*{19pt}
\times \!\!\!\!\prod_{\xi\in {\cal P}_{\alpha\beta}(\gamma)\setminus \gamma}\left\{s_{i\xi}\left[1-\!\!\prod_{\ell\in N_{\xi}(i)}\!\!(1-\sigma_{\ell\xi \to i\xi})\right]\!\right\},\!\!
\label{sum2}
\eea
where 
${\cal P}_{\alpha\beta}(\gamma)$
are all the directed paths of the supernetwork that can be drawn from 
any node $\gamma$  
and that,  starting from the  
superlink $(\gamma,\gamma')$,  arrive at node $\alpha$ from nodes different from  $\beta$.\\

\begin{figure}
\begin{center}
{\includegraphics[width=2.7in]{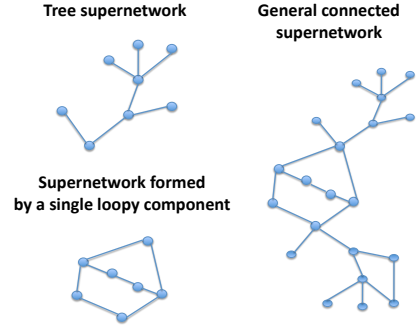}}
\end{center}
\caption{
Schematic view of a several connected supernetwork topologies: a tree supernetwork, a supernetwork formed by a single loopy component and a general connected supernetwork.} 
\label{fig2}
\end{figure}

Let us first consider the case of a tree supernetwork 
(see Fig.~\ref{fig2}). Since the product 
in Eq.~(\ref{sum2}) involves only zeros and ones, we can consider only the paths starting from the leafs of the supernetwork. If $\gamma$ is a leaf of the supernetwork, it follows from Eq.~(\ref{mp1}) that  
$S'_{i\gamma\to i\gamma'}=s_{i\gamma}\left[1-\prod_{\ell\in N_{\gamma}(i)}(1-\sigma_{\ell\gamma \to i\gamma})\right]$.
Therefore we obtain   
 \bea
&&\hspace*{-28pt}
S'_{i\alpha \to i\beta} = \prod_{\xi\in {\cal T}_{\alpha\beta}}\left\{s_{i\xi}\left[1-\prod_{\ell\in N_{\xi}(i)}(1-\sigma_{\ell\xi \to i\xi})\right]\right\},
\label{sumtree}
\eea
where ${\cal T}_{\alpha\beta}$ is the subtree in the supernetwork that has the  root given by the layer $\alpha$ and branching departing from every link of layer $\alpha$ in the supernetwork except the link 
to layer $\beta$.
Using the expression given by Eq.~(\ref{sim}) we 
find for this tree supernetwork that the messages $\sigma_{i\alpha \to j\alpha}$ are given by 
\bea
&&\hspace*{-10pt}
\sigma_{i\alpha \to j\alpha} = 
\prod_{\beta\in 
{\cal G}\setminus \alpha}\left\{s_{i\beta}\left[1-\prod_{\ell\in N_{\beta}(i)}(1-\sigma_{\ell\beta \to i\beta})\right]\right\}
\nonumber \\[5pt]
&&\hspace*{24pt}
\times s_{i\alpha}\left[1-\prod_{\ell\in N_{\alpha}(i)\setminus j}(1-\sigma_{\ell\alpha \to i\alpha})\right],
\label{mestree}
\eea
where we refer the reader to 
Appendix $\ref{ApB}$ for more details. 
Finally, using Eq.~(\ref{S}), it can be shown that  
$S_{i\alpha}$ 
can be 
expressed in terms of the messages as 
\bea
&&\hspace*{-9pt}
S_{i\alpha} = \prod_{\beta\in {\cal G}} 
\left\{\!s_{i\beta}\!\! \left[1-\prod_{\ell\in N_{\beta}(i)}(1-\sigma_{\ell\beta \to i\beta})\right]\!\!\right\}.
\label{Smes}
\eea 
We note here that Eqs.~(\ref{mestree})--(\ref{Smes}) are equivalent to Eqs.~(\ref{mesnew})--(\ref{Smesnew}) on a connected supernetwork where ${\cal C}(\alpha)={\cal G}$ for every $\alpha$. 

Let us prove that Eqs.~(\ref{sum2}) imply that 
Eqs.~(\ref{mestree}) 
are
actually valid 
for every connected supernetwork topology.  
We call the set of layers connected to a 
layer $\alpha$ in the supernetwork at least by two non-overlapping paths 
the {\em loopy component  of layer  $\alpha$}.
We call each connected component formed by  the  layers connected in the supernetwork to layer $\alpha$, but not belonging to the loopy components, the {\em dangling components } of layer $\alpha$. The dangling components might be trees or might contain loops  (see Fig.~\ref{fig2}). Let us assume for simplicity that the 
supernetwork coincides with its loopy component and prove that 
Eq.~(\ref{mestree}) remains valid in this supernetwork topology, which principally differs from trees. 
Since in Eq.~(\ref{sum2})  every path ${\cal P}_{\alpha\beta}(\gamma)$ contributes either by a zero or a one to the product, in order to evaluate the messages $S'_{i\alpha\to i\beta}$, we can consider only the paths ${\cal P}_{\alpha\beta}(\alpha)$ that can be drawn from  layer $\alpha$ and that starting form the 
interlink $(\alpha,\gamma')$ are returning to layer $\alpha$  
through links coming from layers different from  $\beta$. 
Then 
we have 
\bea
&&\hspace*{-28pt}
S'_{i\alpha \to i\beta} = \!\!\!\!\prod_{{\cal P}_{\alpha\beta}(\alpha)} \prod_{\begin{array}{c}\phantom{.}\\[-26pt]\\\scriptstyle{\gamma'}\in{\cal P}_{\alpha\beta}(\alpha)\\[-3pt] \mbox{\scriptsize{with}}\\[-5pt]
\scriptstyle{a_{i\alpha,i\gamma'}=1}\end{array}}\!\!\!\!\!\!\!\!S'_{i\alpha \to i\gamma'}
\nonumber 
\\
&&
\times\prod_{\xi\in {\cal P}_{\alpha\beta}(\alpha)}\left\{s_{i\xi}\left[1-\prod_{\ell\in N_{\xi}(i)}(1-\sigma_{\ell\xi \to i\xi})\right]\right\}
.
\label{sumloopy2}
\eea
This expression is valid for every node $i$ and every pair of interdependent layers $\alpha,\beta$. Since every layer of the loopy component can be reached at least by two non-overlapping paths in the supernetwork, the product over the layers $\xi$ 
in Eq.~(\ref{sumloopy2}) includes all 
layers of the 
supernetwork. 
Therefore, 
all the 
messages 
$S'_{i\alpha \to i\beta}$
of the 
supernetwork ${\cal G}$ 
are all equal to one,    
$S'_{i\alpha \to i\beta}=1$,   
if and only if 
\be
\prod_{\xi\in 
{\cal G}}
\left\{s_{i\xi}\left[1-\prod_{\ell\in N_{\xi}(i)}(1-\sigma_{\ell\xi \to i\xi})\right]\right\}=1
.
\ee
Therefore the messages $S'_{i\alpha \to i\beta}$ can be expressed as 
\bea
&&
S'_{i\alpha \to i\beta} = \prod_{\xi\in 
{\cal G}} 
\left\{s_{i\xi}\left[1-\prod_{\ell\in N_{\xi}(i)}(1-\sigma_{\ell\xi \to i\xi})\right]\right\}.
\label{sumloopy3}
\eea 
Using this result and  Eqs.~(\ref{sim}) and summarizing the results obtained for a supernetwork formed by a loopy component 
we can express the messages 
$\sigma_{i\alpha \to j\alpha}$
in within each layer as 
Eqs.~(\ref{mestree}).

In particular this result 
holds for the special case of a supernetwork formed by a single loop (see Appendix $\ref{ApB}$ for details).
Proceeding in a similar way, 
the result obtained for trees and for loopy components can be directly extended to a connected supernetwork with general topology finding Eqs.~ (\ref{mestree}). 
Indeed, this network is actually formed by a combination of loopy and dangling components, which allows for convenient iterative decomposition and complete analysis of all possible paths of messages  
(see Appendix $\ref{ApC}$ for further details).
Here $S_{i\alpha}$ are given by the same Eq.~(\ref{Smes}). 
Thus 
the mutually connected component  
depends 
only 
on the structure of the 
layers. 
Finally, since every connected component of supernetwork is independent, Eqs.~(\ref{mestree})--(\ref{Smes}) can be generalized for supernetwork topologies with several connected components giving Eqs.~(\ref{mesnew})--(\ref{Smesnew}).

\section{ Average over an ensemble of network of networks with replica nodes}
Let us 
assume that the supernetwork is given (fixed by its adjacency matrix $A_{\alpha\beta}$) and connected and that each 
layer $\alpha$ is generated independently from a configuration model with a degree distribution $P_{\alpha}(k)$. 
That is, in our network of networks, the supernetwork is not random, while the layers are infinite uncorrelated random networks. 
Each layer $\alpha$ has a degree sequence $\{k_i^{\alpha}\}$, and the degrees of the replica nodes in the different layers are uncorrelated. 
Furthermore, 
we  assume that 
nodes  $(i,\alpha)$ are removed from layers with probability $1-p_{\alpha}$ (instigators of cascading failures).
Consequently, the probability  
$P(\{s_{i\alpha}\})$
of the variables 
$s_{i\alpha}$ 
is given by 
\be
P(\{s_{i\alpha}\})=\prod_{\alpha=1}^M\prod_{i=1}^N p_{\alpha}^{s_{i\alpha}}(1-p_{\alpha})^{1-s_{i\alpha}}.
\ee
In order to evaluate  the expected size  of the mutually connected component, 
we 
average the messages over this ensemble of the network of networks.
We indicate with $\sigma_{\alpha}$ the average message in within a layer 
$\Avg{\sigma_{i\alpha \to j\alpha}}=\sigma_{\alpha}$. 
The  equations for the average messages in within a layer are given in terms of the parameters 
$p_{\beta}=\Avg{s_{i\beta}}$
and 
the generating functions $G_0^{\beta}(z)$ and $G_1^{\beta}(z)$ defined as
$G_0^{\beta} = \sum_{k}P_{\beta}(k)z^k$ and 
$G_1^{\beta} = \sum_k kP_{\beta}(k)/\avg{k}_{\beta}z^{k-1}$, 
respectively. 
Using 
Eq.~(\ref{mestree}) 
we find
\be
\sigma_{\alpha} = 
p_{\alpha}\!\!\!\!\prod_{\beta\in 
{\cal G} \setminus \alpha}\!\!\!\!\left\{p_{\beta}[1-G_0^{\beta}(1{-}\sigma_{\beta})]\right\}\![1-G_1^{\alpha}(1{-}\sigma_{\alpha})],
\label{e110}
\ee
Moreover, using Eq.~(\ref{Smes}) we can derive 
the probability for 
$S_{\alpha}=\Avg{S_{i\alpha}}$ 
that a randomly chosen node in layer $\alpha$ 
belongs to the mutually connected component: 
\be
S_{\alpha}=\prod_{\beta\in 
{\cal G}} 
\left\{p_{\beta}[1-G_0^{\beta}(1-\sigma_{\beta})]\right\}.
\label{e120}
\ee
In particular, if all 
layers $\beta$ 
have the same topology and 
equal probabilities $p_{\beta}=p$, then the average messages in within different layers 
are all equal 
$\sigma_{\beta}=\sigma$ and are given by 
\be
\sigma = p^{M}\left[1-G_0(1-\sigma)\right]^{M-1}[1-G_1(1-\sigma)],
\label{e130}
\ee
where $M$ is the cardinality (number of nodes) of the supernetwork. 
Furthermore, the probability 
$S=\Avg{S_{i\alpha}}$ 
that a node in a given layer is in the mutually connected component is given by 
\be
S=\left\{p[1-G_0(1-\sigma)]\right\}^{M}.
\label{e140}
\ee
Thus, as long as the supernetwork is connected,  the resulting mutual component is determined only by the size of the 
supernetwork, and not by its structure. In particular, this supernetwork can be a fully connected graph. Therefore characterization of this model can be reduced to the problem of finding the mutually connected component of a multiplex network \cite{Havlin1,Dorogovstev}. 
Equations~(\ref{e110}) and (\ref{e120}) or Eqs.~(\ref{e130}) and (\ref{e140}) coincide with those for a multiplex network with 
$M$ layers \cite{Havlin1,Son}. 
We emphasize that these equations are identical to, respectively, Eqs.~(29) and (30) or Eqs.~(34) and (35) from Ref.~\cite{Gao:gbhs2012} obtained for a network of networks with random matching of the nodes in interdependent layers (a node in a layer is interconnected with a single random node in an interdependent layer) having a tree supernetwork. Clearly, this network is identical to the particular case of our network of networks with replica nodes and the same tree supernetwork. Indeed, since interlinks in this network do not form loops, one can relabel nodes in individual layers ascribing the same label to each of interlinked nodes, so that the interlinked nodes are actually replicas. 
This is, however, a particular network of considered in the present work. 
Importantly, we have shown that Eqs. (\ref{e110}) and (\ref{e120}) or Eqs.~(\ref{e130}) and (\ref{e140}) are valid to a much wider class of networks with replica nodes having an arbitrary connected supernetwork with any loops. 
Note that if 
$M\to \infty$, then $\sigma=0$ or $1$, where the solution $\sigma=1$ can only be found 
in the case of $p=1$.


\section{ Discussion and conclusions}
Our findings show that a mutually connected component is independent on the structure of a supernetwork for any networks of networks with replica nodes and a fixed supernetwork. We proved that in these networks the 
phenomenon of the independence of the topology is actually not about the presence or absence of loops in the networks of networks but rather about correlations in the set of interdependencies between nodes in pairs of layers. 
In particular, the results obtained above are valid 
for a ring of interdependent networks, i.e., for a supernetwork having a form of a simple loop.

Interestingly,  
our results are still valid even for more general networks of networks than were considered in this paper. Let us return to the direct consequences of the definition of the mutual component, which we discussed introducing the model of a network of networks. 
Consider again a connected network of networks of this type (see Fig.~\ref{f1}).    
Let us rearrange separately and independently interconnections between nodes in each individual set, 
that consists of node $i$ and all its replicas, retaining each of this sets connected. 
Clearly, this rearrangement 
preserves the set of nodes belonging to the mutually connected component. 
On the other hand, after this rearrangement, different nodes in the same layer of the resulting network of networks already can be interconnected with different sets of layers, which differs principally from Fig.~\ref{f1}.

The form of our final results, Eqs.~(\ref{e110}) and (\ref{e120}) or Eqs.~(\ref{e130}) and (\ref{e140}), assumes that each of the individual layers is an uncorrelated network. We presented however simple arguments allowing us to suggest that our qualitative conclusion, namely, the independence on the topology of the supernetwork, is actually valid for arbitrary structures of the layers.   

Finally, we 
should 
indicate an alternative class of networks of networks in which organization of interconnections really matters. 

For example, let the 
interlinks connecting nodes in different layers be completely random similarly to the configuration model. In other words, 
the 
interlinks 
of the nodes of each layer  
are assumed to be uncorrelated in contrast to Fig.~\ref{f1}.  
The detailed general analysis of this class of networks of networks show that  percolation  in these networks of networks  is irreducible to multiplex networks  and that these networks of networks display  multiple percolation transitions \cite{BianDoro}.

In summary we have shown that in a surprisingly wide class of networks of networks, a mutually connected component does not depend on their global topology. The mutual component problem for these networks of networks is resolved by reduction to that for multiplex networks.


\acknowledgments

This work was partially supported by the FCT
projects PTDC: SAU-NEU/103904/2008 and MAT/114515/2009, PEst-C/CTM/LA0025/2011, and FET IP Project MULTIPLEX 317532.




\appendix

\section{Derivation of the simplified expression for the messages $\sigma_{i\alpha\to j\alpha}$ given by Eq.~(2)}
\label{ApA}
In this appendix,  
from the definition in Eqs.~(1)  that we rewrite here for convenience,
\bea
&&\!\!\!\!\!\!\!\!\!\!\!\!\!\!\!\!\!\!
\sigma_{i\alpha\to j\alpha} = s_{i\alpha}\!\!\!\!\prod_{\beta\in {\cal N}(\alpha)}\!\!\!\!S'_{i\beta \to i\alpha}
\!\left[1-\!\!\!\!\prod_{\ell\in N_{\alpha}(i)\setminus j}\!\!\!\!(1-\sigma_{\ell\alpha \to i\alpha})\right],
\nonumber 
\\[5pt]
&&\!\!\!\!\!\!\!\!\!\!\!\!\!\!\!\!\!\!
S'_{i\alpha \to i\beta} = s_{i\alpha}\!\!\!\!\prod_{\gamma\in {\cal N}(\alpha)\setminus \beta}\!\!\!\!S'_{i\gamma \to i\alpha}
\!\left[1-\!\!\!\!\prod_{\ell\in N_{\alpha}(i)}\!\!\!\!(1-\sigma_{\ell\alpha \to i\alpha})\right]
,
\label{mp1A}
\eea
we derive 
 the simplified expression for the messages $\sigma_{i\alpha\to j\alpha}$ given by Eq.~(2), i.e.,
\bea
&&\!\!\!\!\!\!\!\!\!\!\!\!\!\!\!\!\!\!\!
\sigma_{i\alpha \to j\alpha} = S'_{i\alpha\to i\beta}S'_{i\beta \to i\alpha}
\!\left[1-\!\!\!\!\prod_{\ell\in N_{\alpha}(i)\setminus j}\!\!\!\!(1-\sigma_{\ell\alpha \to i\alpha})\right]
.
\label{simA}
\eea

Let us rewrite Eqs.~(\ref{mp1A}), introducing the auxiliary message 
$y_{i\alpha \to j\beta}=s_{i\alpha}\prod_{\gamma\in {\cal N}(\alpha)\setminus\beta}S'_{i\gamma \to i\alpha}$. Then we have 
\bea
&&\!\!\!\!\!\!\!\!\!\!\!
\sigma_{i\alpha\to j\alpha}=y_{i\alpha\to i\beta}\,S'_{i\beta \to i\alpha}
\!\left[1-\!\!\!\!\prod_{\ell\in N_{\alpha}(i)\setminus j}\!\!\!\!(1-\sigma_{\ell\alpha \to i\alpha})\right],
\nonumber 
\\[5pt]
&&\!\!\!\!\!\!\!\!\!\!\!
S'_{i\alpha \to i\beta} = y_{i\alpha \to i\beta}
\left[1-\prod_{\ell\in N_{\alpha}(i)}(1-\sigma_{\ell\alpha \to i\alpha})\right].
\eea
We notice that  
$\sigma_{i\alpha \to j\alpha}=1$ 
only if also 
$\left[1-\prod_{\ell\in N_{\alpha}(i)\setminus j}(1-\sigma_{\ell\alpha \to i\alpha})\right]=1$. 
The condition 
$\left[1-\prod_{\ell\in N_{\alpha}(i)\setminus j}(1-\sigma_{\ell\alpha \to i\alpha})\right]=1$  
implies that automatically 
$\left[1-\prod_{\ell\in N_{\alpha}(i)}(1-\sigma_{\ell\alpha \to i\alpha})\right]=1$.

We summarize these considerations 
writing the following system of equations for the messages 
\bea
&&\!\!\!\!\!\!\!\!\!\!\!\!\!\!\!\!
\sigma_{i\alpha \to j\alpha} = S'_{i\alpha\to i\beta}S'_{i\beta \to i\alpha}
\!\left[1-\!\!\!\!\prod_{\ell\in N_{\alpha}(i)\setminus j}\!\!\!\!(1-\sigma_{\ell\alpha \to i\alpha})\right],
\nonumber 
\\[5pt]
&&\!\!\!\!\!\!\!\!\!\!\!\!\!\!\!\!
S'_{i\alpha \to i\beta} = s_{i\alpha}\!\!\!\!\prod_{\gamma\in {\cal N}(\alpha)\setminus\beta}\!\!\!\!S'_{i\gamma \to i\alpha}
\left[1-\!\!\!\!\prod_{\ell\in N_{\alpha}(i)}\!\!\!\!(1-\sigma_{\ell\alpha \to i\alpha})\right] 
\label{sum1}
\eea
that must be satisfied for every connected pair of layers $(\alpha,\beta)$.


\section{The application of the message passing equations to simple networks }
\label{ApB}

In this appendix
 we 
provide simple examples 
for which we 
verify 
the validity of the general relations given by Eqs.(\ref{mesnew}) and $(\ref{Smesnew})$.
For the sake of convenience, we rewrite here these relations, 
\bea
&&\hspace*{-10pt}
\sigma_{i\alpha \to j\alpha} = 
\prod_{\beta\in {\cal C}({\alpha})\setminus \alpha}\left\{s_{i\beta}\left[1-\prod_{\ell\in N_{\beta}(i)}(1-\sigma_{\ell\beta \to i\beta})\right]\right\}
\nonumber \\[5pt]
&&\hspace*{24pt}
\times s_{i\alpha}\left[1-\prod_{\ell\in N_{\alpha}(i)\setminus j}(1-\sigma_{\ell\alpha \to i\alpha})\right],
\label{mesA}\nonumber \\[5pt]
&&\hspace*{-9pt}
S_{i\alpha} = \prod_{\beta\in {\cal C}(\alpha)} \left\{\!s_{i\beta}\!\! \left[1-\prod_{\ell\in N_{\beta}(i)}(1-\sigma_{\ell\beta \to i\beta})\right]\!\!\right\}
, 
\label{SmesA}
\eea 
where ${\cal C}(\alpha)$ is the connected component of the supernetwork to which the layer $\alpha$ belongs. In particular, if the supernetwork contains only a single connected component, ${\cal C}(\alpha)={\cal G}$ for every layer $\alpha$, and   the cardinality of this component (its size) $|{\cal C}(\alpha)|$ coincides with $M$.


\subsubsection{Message passing equations~(\ref{mp1}) on a forest supernetwork}

Let us 
consider the message passing Eqs.~(\ref{mp1}) on a forest supernetwork, i.e., a supernetwork  formed by several components, in which each component is a tree.
 Starting from the message passing equations 
\bea
\!\!\!\!\!\!\sigma_{i \alpha\to j\alpha}&=&s_{i\alpha}\!\!\!\prod_{\beta\in {\cal N}(\alpha)}\!\!\!S'_{i\beta\to i\alpha}\!\!
\left[1-\!\!\!\prod_{\ell\in N_{\alpha}(i)\setminus j}\!\!\!(1{-}\sigma_{\ell\alpha\to i\alpha})\right]
,
\nonumber 
\\[5pt]
\!\!\!\!\!\!S'_{i\alpha\to i\beta}&=&s_{i,\alpha}\!\!\!\!\!\!\prod_{\gamma\in {\cal N}(\alpha)\setminus \beta}\!\!\!\!\!\!S'_{i,\gamma\to i\alpha}\!\!
\left[1-\!\!\!\!\prod_{\ell\in N_{\alpha}(i)}\!\!\!\!(1{-}\sigma_{\ell\alpha\to i\alpha})\right].
\eea
We notice that the messages can be expressed  as   
\bea
\sigma_{i\alpha\to j\alpha}&=&y_{i\alpha\to i\beta}S'_{i\beta\to i\alpha}
\nonumber \\[5pt]
&&\times\left[1-\prod_{\ell\in N_{\alpha}(i)\setminus j}(1-\sigma_{\ell\alpha\to i\alpha})\right]
,
\nonumber \\[5pt]
S'_{i\alpha\to i\beta}&=&
y_{i\alpha\to i\beta}
\left[1-\prod_{\ell\in N_{\alpha}(i)}(1-\sigma_{\ell \alpha\to i,\alpha})\right]
,
\nonumber \\[5pt]
y_{i\alpha\to i\beta}&=&s_{i\alpha}\prod_{\gamma\in {\cal N}(\alpha)\setminus\beta}S'_{i\gamma \to i\alpha}.
\eea
We observe that  $\sigma_{i\alpha\to j\alpha}=1$ only if also $\left[1-\prod_{\ell\in N_{\alpha}(i)\setminus j}(1-\sigma_{\ell\alpha \to i\alpha})\right]=1$. 
The condition 
$\left[1-\prod_{\ell\in N_{\alpha}(i)\setminus j}(1-\sigma_{\ell \alpha \to i\alpha})\right]=1$  
implies that automatically 
$\left[1-\prod_{\ell\in N_{\alpha}(i)}(1-\sigma_{\ell\alpha\to i\alpha})\right]=1$.

We can summarize these considerations writing the following equations for the messages
\bea
\sigma_{i\alpha\to j\alpha}&=&S'_{i\alpha \to i\beta}S'_{i\beta\to i\alpha}
\nonumber \\[5pt]
&&\times\left[1-\prod_{\ell\in N_{\alpha}(i)\setminus j}(1-\sigma_{\ell \alpha \to i\alpha})\right]
,
\nonumber \\[5pt]
S'_{i\alpha \to i\beta}&=&s_{i,\alpha}\prod_{\gamma\in {\cal N}(\alpha)\setminus\beta}S'_{i\gamma\to i\alpha}
\nonumber \\[5pt]
&&\times \left[1-\prod_{\ell\in N_{\alpha}(i)}(1-\sigma_{\ell\alpha\to i\alpha})\right]
\eea
that must be satisfied for every connected pair of networks $(\alpha,\beta)$.
If we consider a connected network, the last equation for the messages $S'_{i\alpha\to i\beta}$ has the form 
\be
\hspace*{-8pt}
S'_{i\alpha\to i\beta} = \!\!\!\prod_{{\cal P}_{\alpha\beta}(\gamma)}\ \prod_{\xi\in {\cal P}_{\alpha\beta}(\gamma)}\!\!\left\{\!s_{i\xi}\!\left[1-\!\!\!\prod_{\ell\in N_{\xi}(i)}\!\!\!(1{-}\sigma_{\ell\xi\to i\xi})\right]\!\!\right\},
\ee
where ${\cal P}_{\alpha\beta}(\gamma)$ are all the directed paths that can be drawn from any leaf $\gamma$ of 
the 
supernetwork 
and that arrive at node $\alpha$ from nodes different from node $\beta$.
Therefore 
the entire set of  messages $S'_{i\alpha\to i\beta}$ of 
the connected component ${\cal C}(\alpha)$ including 
node (layer) $\alpha$ of a supernetwork, are all equal to one,  $S'_{i\alpha\to i\beta}=1$,  if and only if 
\be
\prod_{\beta\in {\cal C}({\alpha})}\left\{s_{i\beta}\left[1-\prod_{\ell\in N_{\beta}(i)}(1-\sigma_{\ell \beta \to i\beta})\right]\right\}=1
.
\ee
Summarizing the results of this section  we can say that the messages $\sigma_{i\alpha \to j\alpha}$ in within each layer is given by 
\bea
\hspace*{-10pt}\sigma_{i\alpha\to j\alpha}&=&\!\!\!\prod_{\beta\in {\cal C}({\alpha})\setminus \alpha}\!\!\left\{s_{i\beta}\left[1-\!\!\!\prod_{\ell\in N_{\beta}(i)}(1-\sigma_{\ell\beta\to i\beta})\right]\right\}
\nonumber 
\\[5pt]
\hspace*{-20mm}&&\times s_{i\alpha}\left[1-\prod_{\ell\in N_{\alpha}(i)\setminus j}(1-\sigma_{\ell\alpha \to i\alpha})\right]
.
\label{mes2}
\eea
Finally $S_{i\alpha}$ indicates if   a node $(i,\alpha)$ is in the mutually interdependent 
component ($S_{i\alpha}=1,0$), and this indicator function can be expressed in terms of the messages as 
\bea
S_{i\alpha}&=&s_{i\alpha}\prod_{\beta\in {\cal C}(\alpha)\setminus \alpha}  \left[1-\prod_{\ell\in N_{\beta}(i)}(1-\sigma_{\ell\beta\to i\beta})\right]
\nonumber \\
&&\times \left[1-\prod_{\ell\in N_{\alpha}(i)}(1-\sigma_{\ell\alpha \to i\alpha})\right]
.
\eea 


\subsubsection{Message passing equations~(\ref{mp1}) on a single loop supernetwork}

In this subsection we  consider  the message passing equations on a  supernetwork formed by a  loop 
and find similar results.
 We indicate the layers as $\alpha=1,2\ldots M$, where each layer $\alpha$ is linked to the layers $\alpha+1$ and $\alpha-1$. Here we identify layer $M+1$ with layer $1$ and layer $0$ with layer $M$. 
The original message passing equations applied to this simple supernetwork topology are given by 
\bea
\sigma_{i\alpha\to j\alpha}&=&s_{i\alpha}S'_{i\alpha+1\to i\alpha}S'_{i\alpha-1\to i\alpha}
\nonumber \\[5pt]
&&\times\left[1-\prod_{\ell\in N_{\alpha}(i)\setminus j}(1-\sigma_{\ell\alpha\to i\alpha})\right]
,
\nonumber \\[5pt]
S'_{i\alpha\to i\alpha\pm1}&=&s_{i\alpha}S'_{i\alpha\pm 1\to i\alpha}
\nonumber \\[5pt]
&&\times \left[1-\prod_{\ell\in N_{\alpha}(i)}(1-\sigma_{\ell\alpha\to i\alpha})\right].
\label{uno}
\eea
Solving recursively the  equations for $S'_{i\alpha\to i\alpha\pm1}$ we get 
\bea
S'_{i\alpha \to i\alpha\pm1} =S'_{i\alpha\to i\alpha\pm 1}\prod_{\alpha}\!\left\{\!s_{i\alpha}\!\!\left[1- \!\!\!\prod_{\ell\in N_{\alpha}(i)}\!\!\!\!(1{-}\sigma_{\ell\alpha\to i\alpha})\right]\!\!\right\} 
\nonumber
\eea
which yields the solution $S'=S'_{i\alpha\to i\alpha\pm 1}=1$ only if 
\bea
\prod_{\beta=1,M}s_{i\beta}\left[1-\prod_{\ell\in N_{\beta}(i)}(1-\sigma_{\ell\beta\to i\beta})\right]=1,
\eea
i.e., only if all the ``replica'' nodes $(i,\alpha)$ of node $i$  are in the mutually connected component.
Therefore Eqs.~(\ref{uno}) reduce to 
\bea
\hspace*{-5mm}\sigma_{i\alpha\to j\alpha}&=&\prod_{\beta=1,M}\left\{s_{i\beta}\left[1-\prod_{\ell\in N_{\beta}(i)}(1-\sigma_{\ell\beta\to i\beta})\right]\right\}
\nonumber \\[5pt]
\hspace*{-25mm}&&\times s_{i\alpha}\left[1-\prod_{\ell\in N_{\alpha}(i)\setminus j}(1-\sigma_{\ell\alpha\to i\alpha})\right]
.
\eea
Now we notice that 
\bea
&&\hspace*{-5mm}\left[1-\prod_{\ell\in N_{\alpha}(i)}(1-\sigma_{\ell\alpha\to i\alpha})\right]\!\!
\left[1-\prod_{\ell\in N_{\alpha}(i)\setminus j}(1-\sigma_{\ell\alpha\to i\alpha})\right]
\nonumber 
\\[5pt]
&&\hspace*{5mm}=\left[1-\prod_{\ell\in N_{\alpha}(i)\setminus j}(1-\sigma_{\ell\alpha\to i\alpha})\right]
\eea
because if node $(i,\alpha)$ has at least one neighbor $(\ell,\alpha)\neq (j,\alpha)$ in the mutually connected component, then we have automatically 
\bea
\left[1-\prod_{\ell\in N_{\alpha}(i)}(1-\sigma_{\ell \alpha\to i\alpha})\right]=1
.
\eea 
Therefore the equations for the loop supernetwork are given by 
\bea
\hspace*{-20pt}
\sigma_{i\alpha\to j\alpha}&=&\prod_{\beta\neq \alpha}\left\{s_{i\beta}\left[1-\prod_{\ell\in N_{\beta}(i)}(1-\sigma_{\ell\beta\to i\beta})\right]\right\}
\nonumber 
\\[5pt]
&&
\times s_{i\alpha}\left[1-\prod_{\ell\in N_{\alpha}(i)\setminus j}(1-\sigma_{\ell \alpha \to i\alpha})\right]
.
\label{sml}
\eea
Finally $S_{i\alpha}$ indicates if   a node $(i,\alpha)$ is in the mutually connected component ($S_{i\alpha}=1,0$) and  can be expressed in terms of the messages as 
\bea
S_{i\alpha}=s_{i\alpha}\prod_{\beta\neq\alpha}  \left[1-\prod_{\ell\in N_{\beta}(i)}(1-\sigma_{\ell\beta\to i\beta})\right]
\nonumber \\[5pt]
\hspace*{-20mm}\times \left[1-\prod_{\ell\in N_{\alpha}(i)}(1-\sigma_{\ell\alpha \to i\alpha})\right]
.
\eea


\section{Derivation of Eqs.~(\ref{mestree}) and (\ref{Smes})   for a  network of networks with replica nodes and connected supernetwork.}
\label{ApC}

Given a node (layer)  $\alpha$ of the supernetwork,  the set of layers connected to the layer $\alpha$ at least by two non-overlapping  paths,  is called the  {\em loopy component  of layer  $\alpha$} in the supernetwork. We call each connected component formed by  the  layers connected in the supernetwork to layer $\alpha$, but not belonging to the loopy components, the {\em dangling components } of layer $\alpha$. The dangling components might be trees or might contain loops  (see Fig.~2).
In Sec.~\ref{par3} we have shown that for a tree supernetwork topology the messages $S'_{i\alpha\to j\beta}$ are given by  Eq.~(\ref{sumtree}) in the main text, i.e., 
 \bea
&&\hspace*{-28pt}
S'_{i\alpha \to i\beta} = \prod_{\xi\in {\cal T}_{\alpha\beta}}\left\{s_{i\xi}\left[1-\prod_{\ell\in N_{\xi}(i)}(1-\sigma_{\ell\xi \to i\xi})\right]\right\},
\label{sumtreeA}
\eea
where ${\cal T}_{\alpha\beta}$ is the subtree in the supernetwork that has the  root given by the layer $\alpha$ and branching departing from every link of layer $\alpha$ in the supernetwork except the link connected to layer $\beta$.
For a supernetwork topology formed by a single loopy component the messages, instead the messages $S'_{i\alpha\to j\beta}$ have been found to follow Eq.~(\ref{sumloopy3}), i.e.
 \bea
&&\hspace*{-28pt}
S'_{i\alpha \to i\beta} = \prod_{\xi\in {\cal C}(\alpha)
}\left\{s_{i\xi}\left[1-\prod_{\ell\in N_{\xi}(i)}(1-\sigma_{\ell\xi \to i\xi})\right]\right\}.
\label{suA}
\eea
These two expressions [Eqs.~(\ref{sumtreeA}) and (\ref{suA})] are both equivalent to the following one 
 \bea
&&\hspace*{-28pt}
S'_{i\alpha \to i\beta} = \prod_{\xi\in {\cal C}_{\alpha\beta}}\left\{s_{i\xi}\left[1-\prod_{\ell\in N_{\xi}(i)}(1-\sigma_{\ell\xi \to i\xi})\right]\right\},
\label{suGA}
\eea
where ${\cal C}_{\alpha\beta}$ is the connected component of the supernetwork that can be reached from layer $\alpha$ departing from links pointing to layers different from $\beta$. In fact in the case of a tree supernetwork ${\cal C}_{\alpha\beta}={\cal T}_{\alpha\beta}$ and in the case of a loopy component ${\cal C}_{\alpha\beta}={\cal C}(\alpha)$.
From these expressions it follows that for the tree supernetwork and for the supernetwork formed by a single loopy component, the messages  $\sigma_{i\alpha\to j\alpha}$ and the indicator function $S_{i\alpha}$ are given by 
\bea
&&\hspace*{-10pt}
\sigma_{i\alpha \to j\alpha} = 
\prod_{\beta\in {\cal C}({\alpha})\setminus \alpha}\left\{s_{i\beta}\left[1-\prod_{\ell\in N_{\beta}(i)}(1-\sigma_{\ell\beta \to i\beta})\right]\right\}
\nonumber 
\\[5pt]
&&\hspace*{24pt}
\times s_{i\alpha}\left[1-\prod_{\ell\in N_{\alpha}(i)\setminus j}(1-\sigma_{\ell\alpha \to i\alpha})\right],
\label{mesB} 
\\[5pt]
&&\hspace*{-9pt}
S_{i\alpha} = \prod_{\beta\in {\cal C}(\alpha)} \left\{\!s_{i\beta}\!\! \left[1-\prod_{\ell\in N_{\beta}(i)}(1-\sigma_{\ell\beta \to i\beta})\right]\!\!\right\}.
\label{SmesB}
\eea 
Here we 
show that these equations are valid for every supernetwork topology.
Let us consider a completely general supernetwork and a layer $\alpha$ in the supernetwork. The supernetwork can be decomposed into the loopy component of layer $\alpha$ and the dangling components of layer $\alpha$. Given a generic layer $\gamma$ belonging to one of such dangling components we can further decompose this dangling component into the loopy component of layer $\gamma$ and the dangling components of layer $\gamma$. Continuing this iteration we will arrive at the end of the iteration in which we will  
have dangling components with a tree topology.
Given this possible decomposition of the supernetwork, here we prove the validity of Eqs.~(\ref{suGA}), (\ref{mesB}) and Eqs.~(\ref{SmesB}) using an iterative argument.
In particular we want to prove that if  we assume that the messages coming from every dangling components of layer $\alpha$ follow Eqs.~(\ref{suGA}),  then the messages in the loopy component of layer $\alpha$ also follow Eqs.~(\ref{suGA}) and therefore Eqs.~(\ref{mesB}) and (\ref{SmesB}).
Starting from the message passing Eqs.~(\ref{mp1}) valid for $S'_{i\alpha \to i\beta}$, we can follow the messages coming to node $(i,\alpha)$ from other layers different from layer $\beta$ backwards. The paths starting from the layer $\alpha$ and departing from the links of the supernetwork connecting layers different from layer $\beta$, can be distinguished between paths belonging only to the loopy component of layer $\alpha$, that are reaching again layer $\alpha$, closing a loop, or paths that reach the dangling components of layer $\alpha$.
Therefore, if we indicate with ${\cal M}_{\alpha}$ the neighbors of layer $\alpha$ in the loopy component,  by ${\cal D}_{\alpha}$ the layers belonging to  the dangling components of layer $\alpha$ linked to the loopy component of layer $\alpha$,  and finally if we denote by ${\cal L}_{\alpha}$ the loopy component of layer $\alpha$ we obtain the 
following  expression for the messages $S'_{i\alpha\to i\beta}$ departing  from layer $\alpha$,
\bea
&&\hspace*{-28pt}
S'_{i\alpha \to i\beta} = \!\!\!\!\prod_{{\cal P}_{\alpha\beta}(\alpha)}\!\!\! \prod_{\begin{array}{c}\phantom{.}\\[-26pt]\\\scriptstyle{\gamma}\in {\cal P}_{\alpha\beta}(\alpha)\\[-3pt] \mbox{\scriptsize{with}}\\[-5pt]
\scriptstyle{a_{i\alpha,i\gamma}=1}\end{array}}\!\!\!\!\!\!\!\!
S_{i\alpha\to i\gamma} \!\!\!\!\prod_{\begin{array}{c}\phantom{.}\\[-26pt]\\\scriptstyle{\psi\in{\cal D}(\alpha)\setminus \beta}\\[-3pt] \mbox{\scriptsize{with}}\\[-5pt]
\scriptstyle{\phi\in {\cal L}_{\alpha}}\\[-5pt]\mbox{\scriptsize{and}}\ \scriptstyle{a_{i\psi,i\phi}=1}\end{array}}S'_{i\psi\to i\phi}
\nonumber 
\\[5pt]
&&\hspace*{5pt}\times \prod_{\xi\in {\cal L}_{\alpha}}\left\{s_{i\xi}\left[1-\prod_{\ell\in N_{\xi}(i)}(1-\sigma_{\ell\xi \to i\xi})\right]\right\},
\label{sum2A}
\eea
where  $\gamma \in {\cal M}_{\alpha}$ and ${\cal P}_{\alpha\beta}(\alpha)$ are all the directed paths of the loopy component of the supernetwork  that can be drawn from node $\alpha$ and that starting from the 
interlink $(\alpha,\gamma)$ arrive at layer $\alpha$ from layers different from  $\beta$. The expression (\ref{sum2A})  is valid either if $\beta\in {\cal M}(\alpha)$ or if $\beta\in {\cal D}_{\alpha}$.
Now, assuming that all the messages $S'_{i\psi \to i\phi}$ coming from the dangling components satisfy Eq.~(\ref{suGA}), we find that all the messages $S'_{i\alpha\to i\beta}$ have the same value, and that these messages are given by 1  only if 
 \bea
&&\hspace*{-28pt}
1= \prod_{\xi\in {\cal C}_{\alpha\beta}}\left\{s_{i\xi}\left[1-\prod_{\ell\in N_{\xi}(i)}(1-\sigma_{\ell\xi \to i\xi})\right]\right\},
\label{suGAx}
\eea
where ${\cal C}_{\alpha\beta}$ is the connected component of the supernetwork that can be reached from layer $\alpha$ departing from links pointing to layers different from $\beta$. This implies that Eq.~(\ref{suGA}) remains valid in this case.
This result concludes our proof by iteration that   Eq.~(\ref{suGA}) is valid from every topology of the supernetwork.
Finally using Eq.~(\ref{simA})  it can be be shown that this result implies that also Eqs.~(\ref{mesB}) and (\ref{SmesB}) identical to 
Eqs.~(\ref{mestree}) and (\ref{Smes})  are valid for every connected topology of the supernetwork.






\begin{thebibliography}{99}


\bibitem{Dynamics}
A. Barrat, M. Barth\'elemy, and A. Vespignani, 
{\it Dynamical Processes on Complex Networks} (Cambridge University Press, Cambridge, 2008).

\bibitem{crit} S. N. Dorogovtsev, A.~V. Goltsev, and J.~F.~F. Mendes, 
Rev. Mod. Phys. {\bf 80}, 1275 (2008).

\bibitem{Boccaletti_rev}
S. Boccaletti, G. Bianconi, R. Criado, C. I. Del Genio, J. G\'omez-Garde\~nes, M. Romance, I. Sendi\~na-Nadal, Z. Wang, and M. Zanin. 
Physics Reports {\bf 544}, 1 (2014). 

\bibitem{Arenas_rev}
M. Kivel\"a, A.  Arenas, M. Barth\'elemy, J. P. Gleeson, Y. Moreno, and M.~A. Porter,  
 Journal of Complex Networks {\bf 2}, 203 (2014).
 
\bibitem{Havlin1}
S. V. Buldyrev, R. Parshani, G. Paul, H. E. Stanley, and S. Havlin, 
Nature {\bf 464}, 1025 (2010).

\bibitem{PRE}
G. Bianconi, 
Phys. Rev. E {\bf 87}, 062806 (2013).

\bibitem{Havlin2}
R. Parshani, S. V. Buldyrev, and S. Havlin, 
Phys. Rev. Lett. {\bf 105}, 048701 (2010).

\bibitem{Son}
S.-W. Son, G. Bizhani, C. Christensen, P. Grassberger, and M. Paczuski, 
EPL {\bf 97} 16006 (2012).

\bibitem{Dorogovstev}
G. J. Baxter, S. N. Dorogovtsev, A.~V. Goltsev, and J.~F.~F. Mendes, 
Phys. Rev. Lett. {\bf 109}, 248701 (2012). 

\bibitem{Goh}
B. Min, S.~D.~Yi, K.-M.~Lee, and K.-I.~Goh, 
Phys. Rev. E {\bf 89}, 042811 (2014).

\bibitem{Kabashima}
S. Watanabe and Y. Kabashima, 
Phys. Rev. E. {\bf 89}, 012808 (2014).
 
\bibitem{JSTAT}
K. Zhao and G. Bianconi,  
J. Stat. Mech.  P05005 (2013).

\bibitem{Diffusion} 
S. G\'omez, A. D\'iaz-Guilera, J. G\'omez-Garde\~nes, C.~J. P\'erez-Vicente, Y. Moreno, and A. Arenas, 
{Phys. Rev. Lett.} \textbf{110}, 028701 (2013).

\bibitem{Boguna} 
A. Saumell-Mendiola, M. \'A. Serrano, and M. Bogu\~n\'a, 
{Phys. Rev. E} \textbf{86}, 026106 (2012).

\bibitem{dedomenico} 
M. De Domenico, A. Sole, S. Gomez, and A. Arenas, 
PNAS \textbf{111}, 8351 (2014). 

\bibitem{Havlin3}
J. Gao, S.~V. Buldyrev, H.~E. Stanley, and S.~Havlin, 
Nature Phys. {\bf 8}, 40 (2012). 

\bibitem{Gao:gbhs2012}
J. Gao, S. V. Buldyrev, S. Havlin, and H.~E. Stanley, 
Phys. Rev. E {\bf 85}, 066134 (2012). 

\bibitem{Gao:gbhs2011}
J. Gao, S. V. Buldyrev, S. Havlin, and H.~E. Stanley, 
Phys. Rev. Lett. {\bf 107}, 195701 (2011). 

\bibitem{Dong:dtdxzs2013}
G.~Dong, L.~Tian, R.~Du, J.~Xiao, D.~Zhou, and H.~E. Stanley, 
EPL {\bf 102}, 68004 (2013).

\bibitem{Gao:gbsxh2013}
J.~Gao, S.~V. Buldyrev, H.~E. Stanley, X.~Xu, and S.~Havlin, 
Phys. Rev. E \textbf{88}, 062816 (2013).

\bibitem{Dong:dgdtsh2013}
G.~Dong, J.~Gao, R.~Du, L.~Tian, H.~E.~Stanley, and S.~Havlin, 
Phys. Rev. E {\bf 87}, 052804 (2013). 

\bibitem{BianDoro}
G.~Bianconi and S.~N.~Dorogovtsev, 
Phys. Rev. E {\bf 89}, 062814 (2014). 

\bibitem{BD2}
G.~Bianconi and S.~N.~Dorogovtsev, 
arXiv:1411.4160 (2014).

\bibitem{Mezard}
M. Mezard and A. Montanari, 
{\it Information, Physics and Computation}  
(Oxford University Press, Oxford, 2009).

\bibitem{Weigt}
A. K. Hartmann and M.~Weigt, {\it Phase Transitions in Combinatorial Optimization Problems},
(WILEY-VCH, Weinheim, 2005).

\bibitem{karrer2014percolation}
B.~Karrer, M.~E.~J.~Newman, and L.~Zdeborov‡, 
Phys. Rev. Lett. {\bf 113}, 208702 (2014). 

\bibitem{newman2001random}
M.~E.~J.~Newman, S.~H.~Strogatz, and D.~J.~Watts, 
Phys. Rev. E {\bf 64}, 026118 (2001). 
 

\end{thebibliography}
\end{document}